# Differential coverage:
## automating coverage analysis

Henry Cox
Mediatek, USA
Woburn, MA USA
henry.cox@mediatek.com

*Abstract*—While it is easy to automate coverage data collection, it is a time consuming/difficult/expensive manual process to analyze the data so that it can be acted upon. Complexity arises from numerous sources, of which untested or poorly tested legacy code and third-party libraries are two of the most common.

*Differential coverage* and *date binning* are methods of combining coverage data and project/file history to determine if goals have been met and to identify areas of code which should be prioritized. These methods can be applied to any coverage metric which can be associated with a location – statement, function, expression, toggle, *etc.* – and to any language, including both software (C++, Python, *etc.*) and hardware description languages (SystemVerilog, VHDL). The goal of these approaches is to reduce the cost and the barrier to entry of using coverage data analysis in large-scale projects.

The approach is realized in gendiffcov[1], a recently released open-source tool.

I. INTRODUCTION

Code coverage is a widely used quality metric with a long history[1] in both software and hardware projects. Many coverage metrics have been defined, of which the most common are line, function, and expression[2,3]. Coverage is popular because tool support is available, it is easy to measure and it is easy to understand. On the other hand, it is an entirely negative metric in the sense that, if coverage is less than 100%, its meaning is clear: more tests are required. However, if coverage is 100%, then we have no idea what it means. Code which is not covered is definitely not tested. Code which is covered may or may not be adequately tested.

Applying coverage tools to a real project is a difficult, time consuming task which involves a lot of ongoing manual effort.

- If the coverage target is 100%: developers can look at the report and know exactly which parts of the code need more testing in order to reach 100%. The "are we done yet?" question is easily automated.

- If the coverage target is less than 100%: the coverage reports must be reviewed manually after every build or prior to every release, to determine whether more testing is required and which parts of the code to prioritize.

In practice, 100% code coverage is not a realistic goal for most medium- to large-scale projects, primarily due to legacy code which may not be well tested and to third party libraries for which tests may not exist. It is easy to collect coverage data automatically but coverage data analysis and interpretation remains an expensive, mostly manual process – for example, to check whether particular uncovered code is acceptable in the release or not.

*Differential coverage* and *date binning* are related, complementary approaches to automate the coverage review process by combining coverage data with file/project history to categorize the coverage information. Categorization draws attention to sections of code which have changed but are not exercised or which have not changed but are no longer exercised.

The fundamental premise of this analysis is that recent changes are more important than older code. There are two reasons for this belief:

- Understanding: the author of the recent change is likely to be familiar with what it is intended to do, how it is implemented, and how to exercise it. In contrast, the author may no longer recall much about old code – and might no longer be with the company.

- History: unexercised old code has not caused a problem at user sites up to now. The expectation is that it will not cause a problem in the future.

Differential coverage compares two versions of the implementation – the *baseline* and the *current* versions – and the coverage results for each to segment the code into categories. Certain categories are likely to be more interesting than others – for example, it is more important to review new code which is not exercised than pre-existing code which is exercised.

Date binning combines file history from the project revision control system[5,6] with coverage data to highlight areas of unexercised code which warrant additional attention based on the date of the most recent change. It is then simple to distinguish recently added or modified code which has not been tested from older, presumed stable code which is also not tested.

---

[1] https://github.com/henry2cox/lcov/tree/diffcov_initial
A pull request has been initiated, to merge this code back into lcov/master.

Compared to differential coverage, date binning can have a longer view into file history – *i.e.*, older than the most recent release or the release which was used as the baseline – but with reduced categorization fidelity.

When differential coverage and date binning are combined, we gain the benefits of both.

Both differential coverage categorization and date binning can be applied to any metric whose coverpoints are associated with a location – for example, line, expression, branch, function, toggle, *etc.* coverage. When coverpoints have a location, change history for the location can be combined with the coverage data for that location. It is not necessary to analyze the source code or to have any semantic understanding of the code in order to process and display the result of both of these approaches.

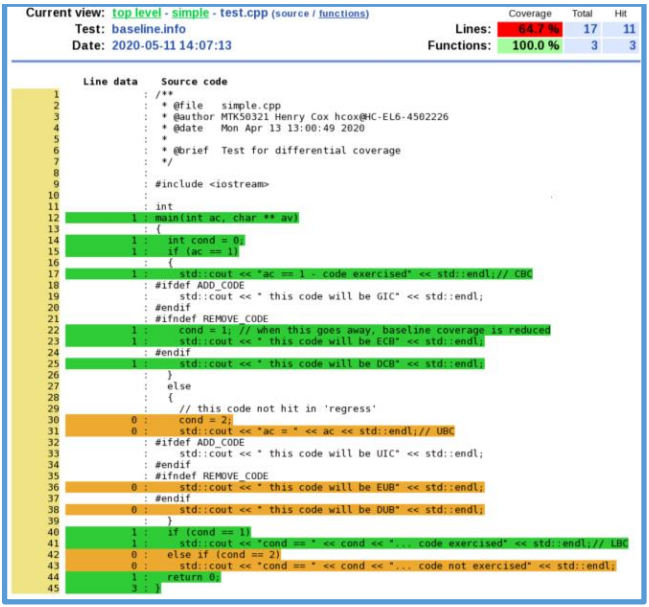

Figure 1 Baseline: g++ --coverage simple.cpp

The remainder of this discussion focuses on C/C++ line coverage – primarily because it is simple and easy to understand, easy to visualize, and because it is familiar to most readers.

## II. DIFFERENTIAL COVERAGE CATEGORIES

Figure 1 and Figure 2 [2] display line coverage for the 'baseline' and 'current' versions of a sample program that is used throughout this discussion. A few lines are inserted (*e.g.*, lines 9 and 15 of Figure 2 – the latter containing code and the former a blank line), a few are deleted (*e.g.*, line 25 of Figure 1), and the compilation command is changed between the revisions. A coverage trace file is generated by executing each *a.out* once, with no arguments. The result of these (admittedly contrived) changes is to create at least one line in each differential category.

The input to the differential coverage process is the baseline coverage result, the current coverage result and the universal diff of the source code file revisions. Our implementation uses LCOV[4] *.info* files to hold coverage results, and uses our revision control systems[5,6] to produce the difference report. We keep track of "inserted lines" and "deleted lines"; we do not attempt to track "changed lines". "Changed line" appears as a "deleted line"/"inserted line" pair at the same location.

Differential coverage categorizes coverpoints by combining answers to two questions:

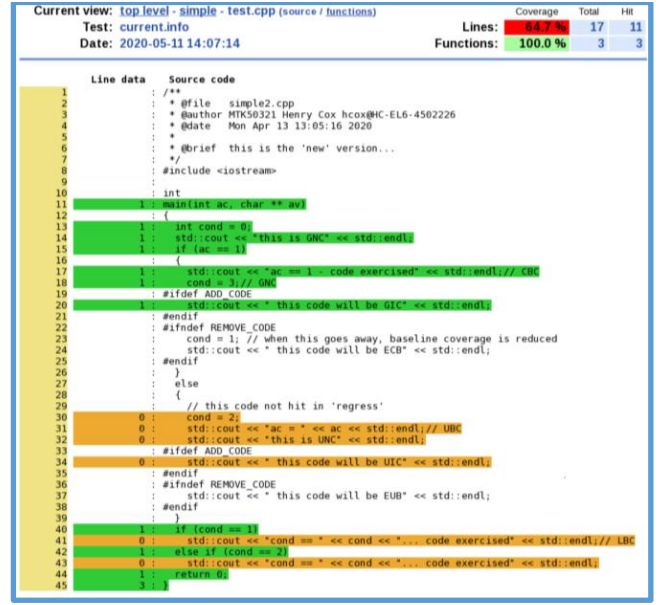

Figure 2 Current: g++ --coverage –DADD_CODE –DREMOVE_CODE simple.cpp

- Is the code present or not present in the baseline and current snapshots?
- Is the code covered or not covered in the baseline and current snapshots?

|  | Covered in current | Not covered in current |
|---|---|---|
| **Covered in baseline** | CBC | LBC |
| **Not covered in baseline** | GBC | UBC |

*Table 1* Code present in both baseline and current snapshots

Table 1 describes the case when the source text is present in both snapshots, and is recognized as code in both snapshots. In all cases, coverage is reported with respect to text from the current version (Figure *2*).

There are four possible coverage combinations:

- **CBC**: "Covered Baseline Code" is covered in both the baseline and current snapshots (no change) – for example, line 13.

---

[2] Figures generated using *LCOV*[4]. Green indicates 'covered' lines and amber uncovered lines. The number in the 'line data' column indicates the number of times the corresponding line was hit; no number means that this line was not recognized as code (*i.e.*, comment, blank line, unused template, *etc.*).

- **GBC**: "Gained Baseline Coverage" is not covered in the baseline but *is* covered in the current snapshot (line 42). The GBC category represents progress: either tests have been added or control paths have changed such that pre-existing code is exercised.
- **LBC**: "Lost Baseline Coverage" is covered in the baseline but *is not* covered in the current snapshot (line 41). This category represents degradation: either tests have been removed or control paths changed such that this code is no longer exercised. A review may indicate that the code is now dead and should be removed, that new tests need to be added, or that a bug was introduced which disabled the code.
- **UBC**: "Uncovered Baseline Code" is not covered in either the baseline or current snapshots (line 30). This situation is not desirable – but it is unchanged from the previous result.

|  | Covered in current | Not covered in current |
|---|---|---|
| **Source text added** | GNC | UNC |
| **Source text unchanged** | GIC | UIC |

*Table 2* Code present in the current snapshot but not present in the baseline (new code)

Table 2 illustrates the case when code is present in the current snapshot but is not present in the baseline. New code can appear in two ways: a developer may have written new code (the baseline source text is changed by the addition of new code) or the compilation conditions may be different such that unchanged source text is identified as code. Amongst the ways that unchanged text can become code are changed conditional compilation directives, removal of a block comment around the text, or changed instantiation of templates and inline functions.

The four possible combinations are:

- **GNC**: "Gained coverage New Code": newly added code is exercised (line 14).
- **GIC**: "Gained coverage of Included Code": unchanged source text is now recognized as code and is exercised by the test suite (line 20). This case is distinct from GNC because the source text of the GIC category has not changed between revisions.
- **UNC**: "Uncovered New Code": newly added code is not exercised (line 32)-
- **UIC**: "Uncovered Included Code": unchanged source text is now recognized as code but is not exercised by the test suite (line 34).

|  | Covered in baseline | Not covered in baseline |
|---|---|---|
| **Source text deleted** | DCB | DUB |
| **Source text unchanged** | ECB | EUB |

*Table 3* Code present in the baseline but not present in the current snapshot (code removed)

Table 3 illustrates the final case: code was present in the baseline but is not present in the current snapshot. The removed code may have been deleted, or compilation conditions may have changed so that unchanged source text is no longer treated as code. Excluded code is categorized differently than deleted code because excluded code is be expected to come back when compilation conditions change again whereas deleted text is unlikely to be restored. As in the previous tables, there are four possible combinations:

- **DCB**: "Deleted Covered Baseline" code had been exercised, but is now deleted (line 25 from Figure 1). By definition deleted code is not present – so it does not appear in line counts, etc. in the current version.
- **ECB**: "Excluded Covered Baseline" code had been exercised; the source text is unchanged, but is no longer recognized as code (*e.g.*, line 23 from Figure 2, within a disabled #ifdef).
- **DUB**: "Deleted Uncovered Baseline" code was not exercised, and is now deleted (line 38 from Figure 1).
- **EUB**: "Excluded Uncovered Baseline" code was not exercised and is no longer recognized as code (line 37 from Figure 2).

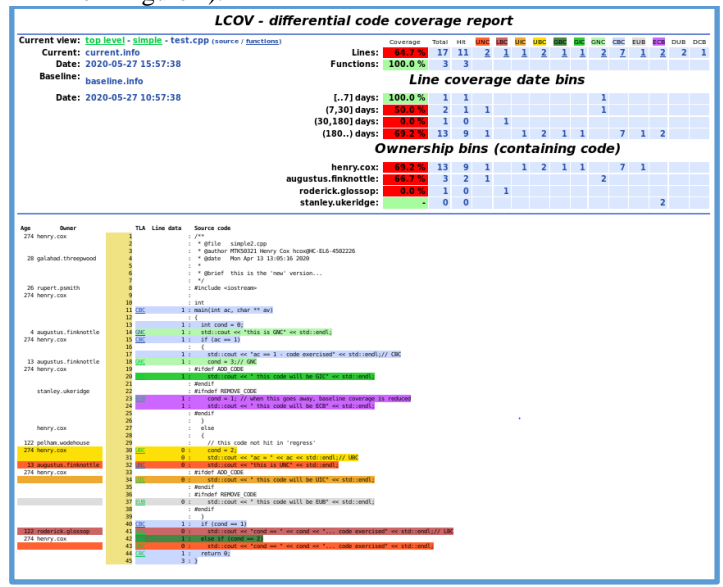

*Figure 3* Differential coverage 'file' view with date- and owner- binning

Table 4 lists the differential coverage categories in priority order. The general principles are that unexercised code is more interesting than exercised code and new code is more interesting than old code. Most developers are unlikely to care about excluded or deleted code – but some will want to review exclusions and deletions to be sure that the changes were expected and were not due to a dropped test or a control path bug which has been introduced.

A property of the differential coverage categories is that there is a one-to-one mapping between them if the sense of "current" and "baseline" is reversed. For example, "GBC" code found when comparing version **B** to version **C** becomes "LBC" code when comparing version **C** to version **B**.

| Acronym | Long name | Meaning |
|---|---|---|
| UNC | Uncovered New Code | Newly added code is not tested |
| LBC | Lost Baseline Coverage | Preexisting code is no longer tested |
| UIC | Uncovered Included Code | Previously unused code is not covered |
| UBC | Uncovered Baseline Code | Preexisting code was not covered before, not covered now |
| GBC | Gained Baseline Coverage | Unchanged code is covered now |
| GIC | Gained coverage Included Code | Previously unused code is covered now |
| GNC | Gained coverage New Code | Newly added code is exercised |
| CBC | Covered Baseline Code | Unchanged code was covered before and is still covered |
| EUB | Excluded Uncovered Baseline code | Previously un-exercised code is unused now. |
| ECB | Excluded Covered Baseline code | Previously exercised code is unused now. |
| DUB | Deleted Uncovered Baseline code | Previously un-exercised code has been deleted. |
| DCB | Deleted Covered Baseline code | Previously exercised code has been deleted. |

Table 4 Differential coverage categories

III. DATE BINNING

Date binning uses file history from the project revision control system to correlate the most recent edit to each line of source code to coverage information tagged to that location – *e.g.*, statement coverage of the corresponding line, expression coverage of the expression found on the line, *etc. Age of last edit* is used to group the coverage data – for example "changed in last 7 days", "changed between 7 and 30 days ago", "changed more than 30 days but less than 6 months ago", and "changed more than 6 months ago". Date intervals should be chosen based on project and release schedules – for example, to catch current development, the previous patch, and the most recent major release. After code in each coverage category is sorted into the appropriate bin, it is trivial to find recently modified source code which is not exercised – as well as the author of the most recent edit of that code.

When differential coverage and date binning are used together, the advantages of both are realized. Some conclusions overlap – for example, the UNC bin will contain the same lines as the 'not covered' date bin with the same age as the baseline. In other cases, the combined information can lead to new insight: LBC code in the latest build may be the responsibility of the author(s) of the commit which triggered the build that saw the coverage loss, or may be the responsibility of the author of the LBC code; both should be notified by the continuous integration platform.

Compared to differential coverage, date binning does not require a baseline and hence can be applied when a baseline is not available. Thus, date binning is a good solution to the *cold start problem.* Date binning is also not subject to the 'broken rachet' effect (see section IV) – as unexercised code continues to appear in the corresponding date bin, even after a new baseline is adopted.

Conversely, without a baseline, date binning cannot identify "lost coverage", nor can it identify excluded or deleted code[3]. In the absence of a baseline, our tool categorizes all covered code as GNC and all uncovered code as UNC. Within these categories, each line is grouped into the appropriate age bin.

Figure 3 shows a combined differential/date binning line coverage report generated by *gendiffcov* when both baseline and file history data are available. Lines of code in each differential category are colorized and the far left column shows the age (in days) and the account who checked in the most recent edit to the corresponding line. Summary tables are generated at the file, directory, and project levels. The screenshot also shows the shows summary table that appears when the "--show-owners" option is specified. The table contains only developers who are associated with executable code (*i.e.*, exclusive of comments, blank lines, *etc.*); with different options, only developers who are responsible for untested code appear in the table[4]. Various navigation features intended to make it easier to understand the current state of the project and what to do next have been implemented – *e.g.*, to navigate to the first line of code in the file in a particular differential category, the next line 'owned' by a particular user in a particular category, *etc.* Such navigation features enable faster analysis and review of the coverage result.

*gendiffcov* is most informative when both baseline and file history are available, but it can also be used with partial data. If file history is not available, the only differential coverage is reported – the left column, the date bin summary, and the owner table will not appear. If baseline data is not available, all code will be categorized "GNC" or "UNC" (hit or not hit). If neither file history nor baseline is available, then only 'traditional' coverage is reported (*i.e.*, as in Figure *2*).

IV. COVERAGE METHODOLOGY

Differential coverage requires a baseline coverage snapshot from some point in the past – *e.g.*, the previous release or the final firmware version from the previous IP generation. However, when coverage is first applied to a new project, the source code and/or build infrastructure may need to be enhanced to enable coverage data collection; it is not possible to go back to generate coverage data for historical releases because those versions do not contain the required infrastructure.

---

[3] Note that the date associated with LBC code is when the source text was last edited – not the date when the coverage loss occurred. To find the date of coverage loss, it is necessary to post-process the continuous integration system logs. This process is outside the scope of this paper.

[4] This is useful in the case of a project with a large number of developers: even though 200 developers may have edited some particular file, the summary table should contain zero (100% coverage) or a few names. If all 200 developers are responsible for untested code – then the project has bigger problems than merely having to wade through very long result tables.

Because it does not require a baseline, date binning offers a path out of the cold-start dilemma – enabling the project team to look back into recent history to improve missed coverage opportunities. Alternately, the project can apply the *New Year's Resolution Approach* – and declare that "today is the baseline…from now on, all commits must satisfy LBC + UNC + UIC == 0".

Differential categorization (UNC + UIC + LBC == 0) can be used as a pass/stable condition in automated regression or continuous integration systems to enable a 'coverage ratchet' which does not allow coverage to decrease. This is a much stronger criteria than simple numeric comparison: for example, line coverage in both Figure *1* and Figure 2 is 64.7% (11 of 17 lines) – overall line coverage is identical, despite that the changes are largely untested.

Care must be taken when choosing to update the baseline used in automated regression or continuous integration systems in order to avoid the *broken ratchet* effect. On one hand, if the baseline is not updated when a new test with non-zero GBC to be added, it is possible for a second change to inadvertently reduce GBC again; the result is a transiently increased coverage which is lost again before it is captured – there is no evidence that it ever happened. On the other hand, if the baseline is updated after code changes whose UNC or LBC is non-zero, then that code will appear as UBC in subsequent builds – and the differential report will no longer draw attention to it. In this case, date binning will continue to show the UNC code as 'recently changed/uncovered' – but the knowledge that LBC category had been covered will be permanently lost. Systematic solutions to these problems exist[7] – but are outside the scope of this discussion.

Coverage analysis tools also need the ability to exclude certain code from analysis. For example, some code is not reachable in normal conditions. Testbenches often contain "cannot happen" error handling code of the form:

```
if (some_failure_happened)
   print_failure_and_die();
```

such that the 'if' clause is reached only if there is a bug in the implementation. Similarly, interrupt handlers and logging interfaces may interact with hardware features which are not available in the simulation framework – and so must be excluded. Unreachable code typically must be analyzed manually then excluded so that it is not reported again[5]. It is important for the regression system to monitor such excluded code as any test which reaches it must be marked as a failure; this prevents the scenario in which the calling code is changed such that formerly unreachable states become reachable – by design or by accident.

Different users may want to use differential coverage data for different purposes. For example, the program manager may want to compare the previous release to the next release, to verify that all new development has been tested (LBC + UNC + UIC == 0). A particular developer may want to verify coverage of their sandbox changes, from branch creation to head, before pushing back to the project repository.

In addition, practical development environments will very likely need coverage signoff process whereby the program manager or project lead can declare that the release should occur despite that the differential coverage criteria or date-bin metric is not met.

V. CONCLUSIONS

Code coverage is a useful quality metric for software and hardware development – but it is easily abused and is difficult, time consuming, and expensive to apply in practice.

A common approach to coverage analysis is to set an arbitrary coverage target – say, 97% – and then stop adding tests when the target is reached. This is a bad idea for many reasons – not least of which is that all code in the release is not created equal. It is very likely that we care more about recently written or changed code than we do about older, stable code.

Differential coverage and date binning are approaches which combine project and file history with coverage data to identify untested regions of code which were recently written or recently changed, as well as regions of unchanged source code which are no longer exercised. This reduces that cost of using code coverage in medium to large scale software or hardware projects, and enables automated analysis such that unexercised code has to be manually analyzed only once – and then that effort amortized over all subsequent builds.

*gendiffcov* is an open-source tool which implements the approaches described above. It has been used to automate coverage data analysis at Mediatek for several software and firmware projects, and is currently being applied to hardware.

ACKNOWLEDGMENTS

I would like to thank Steven Dovich for help with an early implementation of differential coverage, and for helping to integrate the tool into our Jenkins environment. I would also like to thank multiple Mediatek teams for their willingness to try something new.

---

[5] The lcov implementation supports exclusion of particular directories or files. In cases that it is not desirable to exclude an entire file, lcov also supports exclusions via directives (pragmas) in the code. Pragmas are convenient in that they are easy to keep up-to-date when the code changes.

[7] https://www.jenkins.io/